\documentclass{sig-alternate}

\permission{}
\copyrightetc{}

\usepackage{graphicx}
\usepackage{url}
\usepackage[utf8]{inputenc}
\usepackage{psfrag}
\usepackage{subfigure}
\usepackage{hyperref}
\hypersetup{
    colorlinks,%
    citecolor=black,%
    filecolor=black,%
    linkcolor=black,%
    urlcolor=blue
}

\def\sharedaffiliation{%
\end{tabular}
\begin{tabular}{c}}

\usepackage{graphicx}
\begin{document}
%\conferenceinfo{WikiSym}{'2012  Linz Austria}

\title{Biographical Social Networks on Wikipedia} % \\ \Large{Links that made history}}
\subtitle{A cross-cultural study of links that made history}

\numberofauthors{2} %  in this sample file, there are a *total*
\author{
% 1st. author
\alignauthor Pablo Aragon%\\
  %\email{\normalsize{pablo.aragon@barcelonamedia.org}}
% 2nd. author
\alignauthor Andreas Kaltenbrunner%\\
  %\email{\normalsize{andreas.kaltenbrunner@barcelonamedia.org}}
\and
% 3rd. author
  \alignauthor David Laniado
  %\email{\normalsize{david.laniado@barcelonamedia.org}}
% 4th. author
  \alignauthor Yana Volkovich
  %\email{\normalsize{yana.volkovich@barcelonamedia.org}}
%\and
\vspace{4pt}
\sharedaffiliation
  \affaddr{Barcelona Media Foundation, ~ Barcelona, Spain} \\
       %\affaddr{Barcelona, Spain}
\email{\normalsize{\{name.surname\}@barcelonamedia.org}}
}

\maketitle
\begin{abstract}
  It is arguable whether history is made by great men and women or
  vice versa, but undoubtably social connections shape history.
  Analysing Wikipedia, a global collective memory place, we aim to
  understand how social links are recorded across cultures. Starting
  with the set of biographies in the English Wikipedia we focus on the
  networks of links between these biographical articles on the 15
  largest language Wikipedias. We detect the most central characters
  in these networks and point out culture-related peculiarities.
  Furthermore, we reveal remarkable similarities between distinct
  groups of language Wikipedias and highlight the shared knowledge
  about connections between persons across cultures.  
  \sloppy{
  
  }
\end{abstract}

%\iffalse
% \category{H.5.3}{Information Interfaces}{Group and Organisation
%   Interfaces}[Computer-supported cooperative work,Web-based
% interaction] %CHANGE

%\terms{Measurement, Human Factors}

%\fi

 % \vspace{2mm}
%  \noindent {\bf Categories and Subject Descriptors:} H.5.3 [{\bf
%    Information Interfaces}]: Group and Organisation Interfaces -- {\em
%    Computer supported cooperative work, Web-based interaction}

%  \vspace{2mm}
%  \noindent {\bf General Terms:} Human Factors, Measurement

% \vspace{2mm}
% \noindent {\bf Keywords:} Wikipedia, social network analysis, cross
% language studies

\section{Introduction}\label{sec:intro}
Social network analysis, one of the most studied subjects in the last
decade, has been applied in very different contexts ranging from
online social networks~\cite{Mislove2007}, over networks of fictitious
comic characters~\cite{Alberich2002} to animal social
networks~\cite{Chiyo2012}. Here we present a study about %social
connections of a different form. We use neither self-reported nor observed
relations nor interactions inferred from activity logs. We focus on
the %social
links between notable humans as they are recorded in collective
memory. To extract these connections and build the corresponding
networks we use different language versions of Wikipedia, which can be
seen as \emph{global memory place}~\cite{pentzold2009fixing}.

We exploit direct links between biographic articles as evidence of
relations between the corresponding persons, and build biographical
social networks for the 15 largest language versions of Wikipedia.  We
investigate these networks separately and analyse their similarity. We
furthermore extract the most important persons according to several
centrality
% influence
metrics in these networks.  This allows us to analyse and compare the
different language communities over their perception and reporting
of %social
connections between notable persons. The visualisation of the shared
links present in most of the different language networks highlights
the connections commonly known across language and culture barriers.

\section{Related work}\label{sec:related_work}

Social networks analysis on Wikipedia has mainly exploited
editor interactions, either via generating co-authorship
networks~\cite{Laniado11coauthorship} or analysing social interactions
on article and user talk-pages~\cite{Laniado2011}.  Additionally,
%Complementarily,
co-authorship has been used to create networks of similar
articles~\cite{biuk2006visualizing}.
\sloppy{
}

The link structure of Wikipedia articles has been studied extensively:
common features have been found in the network topology of several
language versions of Wikipedia~\cite{zlatic2006wikipedias}, and
rankings of the most central entries in the English Wikipedia have
been presented~\cite{bellomi2005network}.  The idea of restricting the
network to articles representing entities of a given type has been
followed in~\cite{athenikos2009wikiphil}, introducing a framework for
visualising links between philosophers in the English Wikipedia.

Different language versions of Wikipedia have been compared to study
cultural differences among their communities~\cite{pfeil2006cultural}.
While this has been done mostly by analysing the behaviour of the
editors, here we propose to study differences and similarities as they
emerge from the link structure of the artifacts created by different
communities.

\section{Data extraction}\label{sec:methods}

To obtain a list of articles about persons on the English Wikipedia,
we relied on a dataset from
DBpedia\footnote{\scriptsize{\url{http://downloads.dbpedia.org/3.7/en/persondata_en.nt.bz2}}}.
The dataset contains links to 296~511 existing English articles, which
we parsed to identify the names of the corresponding articles in the
other 14 language versions of Wikipedia with the largest number of
articles at the moment we extracted the data (September 8 to 13,
2011).
 
For each language version we generated a directed network where nodes
represent persons with a biographical Wikipedia article and a node $i$
links to node $j$ if the article of the person $i$ is linking to the
article about person $j$.

\begin{figure}[!tb]
%\centering
\includegraphics[width=0.45\textwidth]{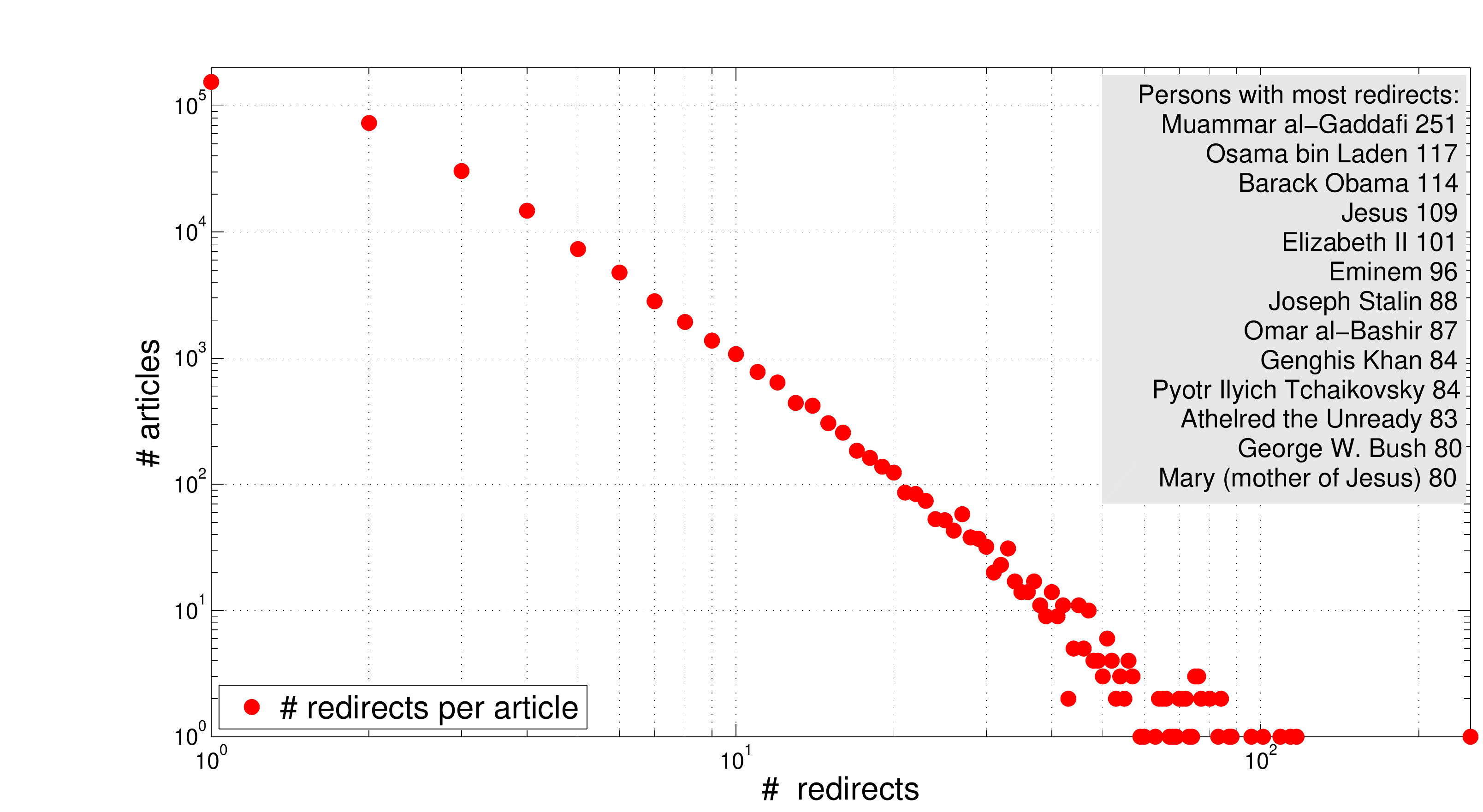}
\caption{Distribution of the number of redirects per biographical article for the
  English Wikipedia.}
\label{fig:redirects_stats}
\end{figure}

As many Wikipedia articles have alternative names which redirect to
the same article, we had to track these redirects for every person and
every language with a script provided by the Wikimedia  
Toolserver.\footnote{\scriptsize{\url{http://toolserver.org/~dispenser/sources/rdcheck.py}}}
The number of redirects per page follows a heavy tailed
distribution as can be observed in 
Figure~\ref{fig:redirects_stats}. The table embedded in the figure
lists the 13 persons with the largest number of redirects in the
English Wikipedia. The article about Muammar al-Gaddafi leads the
ranking with 251 different ways of linking towards it, more than
doubling the redirects of Osama bin Laden, the second ranked person.

\section{Results}\label{sec:results}
In this section we study global metrics calculated for the
biographical networks of different languages. We also discuss rankings
based on various definitions of centrality. In particular, we present
the most central (linked) persons in the different language
Wikipedia. Finally, we compare the similarities between the different
language networks.

\subsection{Global network statistics}\label{sec:hist_network_stats}
A brief overview of the principal social network measures for the
different language networks is given in Table~\ref{tab:NW_measures}.
The largest network corresponds to the English Wikipedia with
nearly $200,000$ nodes. The second largest, extracted from the German
Wikipedia, is only about one-third as large.

All language networks show very low clustering. The only outstanding
network is the Chinese with a clustering coefficient of $0.17$, which
indicates an important structural difference of the link structure in
this language version.

By looking at the link reciprocities we find that it is quite rare
that two persons are mutually connected. One of the possible causes of
this observation may be the presence of parasocial
interactions~\cite{horton1956mass}, i.e. one-sided interpersonal
relationships in which one part knows a great deal about the other,
but the other does not. E.g., when a person is influenced by the works
of somebody who died decades before.

We see that all networks are well connected, as the percentage of
nodes in the giant component (GC)\footnote{The GC
    corresponds here to the largest weakly connected component. {\it
      Weakly connected} means that there exists at least a path in one
    direction between any pair of nodes.}
lies between 85\% (Polish) and 96\% (French and
Japanese).

When calculating the average path length between two persons in the
$GC$ we observe that the largest average distances are found for
the Polish and Russian networks.

\begin{table}[!tb]
  \caption{ Properties of the language networks ordered by
    network size: number of (not isolated) nodes $N$ and edges $K$, average
    clustering coefficient $\langle C \rangle$, percentage of nodes
    in the giant component GC, average path-length between
    nodes $\langle d \rangle$, reciprocity $r$ and maximal distance
    $d_{max}$ between two nodes in the network.}
  \label{tab:NW_measures}
  \centering
  \small
  \begin{tabular}{@{}crrrrccc@{}}\hline
    lang & $N$~~ & $K$~~ & $\langle C \rangle$ &
    \% GC &  $\langle d \rangle$ & $r$ & $d_{max}$\\
    \hline
    en & 198\;190 & 928\;339 & 0.03& 95\% & 6.53 & 0.17 & 43\\
    de & 62\;402 & 260\;889 & 0.05 & 94\% & 6.83 & 0.14 & 33\\
    fr & 51\;811 & 283\;453 & 0.06 & 96\% & 6.11 & 0.15 & 36\\
    it & 35\;756 & 190\;867 & 0.06 & 95\% & 6.28 & 0.14 & 42\\
    es & 34\;828 & 169\;302 & 0.06 & 97\% & 6.29 & 0.16 & 36\\
    ja & 26\;155 & 109\;081 & 0.08 & 96\% & 6.47 & 0.20 & 26\\
    nl & 24\;496 & 76\;651 & 0.08 & 94\% & 7.91 & 0.18 & 37\\
    pt & 23\;705 & 85\;295 & 0.07 & 94\% & 6.98 & 0.18 & 45\\
    sv & 23\;085 & 60\;745 & 0.07 & 91\% & 8.27 & 0.20 & 46\\
    pl & 22\;438 & 50\;050 & 0.08 & 85\% & 8.94 & 0.16 & 43\\
    fi & 18\;594 & 44\;941 & 0.07 & 87\% & 7.80 & 0.17 & 30\\
    no & 18\;423 & 49\;303 & 0.09 & 83\% & 8.31 & 0.22 & 48\\
    ru & 16\;403 & 34\;436 & 0.06 & 87\% & 9.10 & 0.10 & 35\\
    zh & 11\;715 & 44\;739 & 0.17 & 91\% & 7.20 & 0.20 & 32\\
    ca & 11\;027 & 42\;321 & 0.09 & 93\% & 7.14 & 0.17 & 32\\
    \hline
  \end{tabular}
\end{table}

Finally, we also analyse the in- and out-degree distributions and
observe heavy-tails, as found in many real-life
networks, for all language Wikipedias (data not shown).

\begin{table}[!tb]
  \caption{The top 25 persons in the English Wikipedia ranked by
    in-degree. Ranks for out-degree, betweenness and PageRank in
    parenthesis.}
  \label{tab:top_indegree}
  \centering
\scriptsize{
\begin{tabular}{@{}l|r|r@{\hspace{4pt}}r|r|c@{\hspace{6pt}}c@{}}
\hline
    person & in- & \multicolumn{2}{|c|}{out-degree} & btw. & \multicolumn{2}{|c}{PageRank} \\
    \hline 
George W. Bush & 2123 & 89 & (107) & (1) & 0.00209 & (1)\\
  Barack Obama & 1677 & 51 & (710) & (8) & 0.00162 & (2)\\
  Bill Clinton & 1660 & 74 & (205) & (4) & 0.00156 & (4)\\
  Ronald Reagan & 1652 & 90 & (103) & (2) & 0.00156 & (3)\\
  Adolf Hitler & 1407 & 119 & (26) & (3) & 0.00149 & (5)\\
  Richard Nixon & 1299 & 86 & (127) & (7) & 0.00136 & (6)\\
  William Shakespeare & 1229  & 25 & (4203) & (63) & 0.00113 & (9)\\
  John F. Kennedy & 1208 & 104 & (53) & (5) & 0.00123 & (8)\\
  Franklin D. Roosevelt & 1052 & 71 & (237) & (15) & 0.00131 & (7)\\
  Lyndon B. Johnson & 1000 & 106 & (50) & (12) & 0.00108 & (11)\\
  Jimmy Carter & 953 & 80 & (158) & (9) & 0.00113 & (10)\\
  Elvis Presley & 948 & 82 & (142) & (27) & 0.00063 & (24)\\
  Pope John Paul II & 941 & 59 & (444) & (11) & 0.00083 & (18)\\
  Dwight D. Eisenhower & 891 & 55 & (564) & (22) & 0.00095 & (14)\\
  Frank Sinatra & 882 & 108 & (47) & (18) & 0.00056 & (28)\\
  George H. W. Bush & 878 & 87 & (118) & (19) & 0.00096 & (13)\\
  Abraham Lincoln & 846 & 54 & (593) & (40) & 0.00089 & (16)\\
  Bob Dylan & 835 & 151 & (11) & (14) & 0.00055 & (30)\\
  Winston Churchill & 748 & 84 & (136) & (10) & 0.00092 & (15)\\
  Harry S. Truman & 743 & 81 & (145) & (24) & 0.00099 & (12)\\
  Joseph Stalin & 723 & 69 & (265) & (43) & 0.00089 & (17)\\
  Michael Jackson & 663 & 71 & (237) & (34) & 0.00042 & (51)\\
  Elizabeth II & 653 & 52 & (665) & (6) & 0.00074 & (19)\\
  Jesus & 572 & 38 & (1595) & (51) & 0.00068 & (20)\\
  Hillary Rodham Clinton & 554 & 87 & (118) & (32) & 0.00063 & (25)\\
\hline
\end{tabular}}
\end{table}

\begin{table*}[!tb]
  \caption{Top 5 most central persons in the 15 analysed language
    versions of Wikipedia ranked by betweenness.}
  \label{tab:top_betweenness}
   \scriptsize
  \centering
% \tiny
\begin{tabular}{lccccc}
  \hline
  lang & \#{}1 & \#{}2 & \#{}3 & \#{}4 & \#{}5 \\
  \hline
  \textbf{en}& George W. Bush & Ronald Reagan & Adolf Hitler & Bill Clinton & John F. Kennedy\\
  \textbf{de}& Adolf Hitler & George W. Bush & Martin Luther King, Jr & Barack Obama & Frank Sinatra\\
  \textbf{fr}& Adolf Hitler & George W. Bush & William Shakespeare & Barack Obama & Jacques Chirac\\
  \textbf{it}& Frank Sinatra & George W. Bush & Pope John Paul II & Michael Jackson & Elton John\\
  \textbf{es}& Michael Jackson & Fidel Castro & William Shakespeare & Che Guevara & Adolf Hitler\\
  \textbf{ja} & Adolf Hitler & Michael Jackson & Ronald Reagan & Yukio Mishima & Barack Obama\\
  \textbf{nl}& Elvis Presley & Adolf Hitler & Bill Clinton & Joseph Stalin & William Shakespeare\\
  \textbf{pt}& Michael Jackson & Richard Wagner & Adolf Hitler & Ronald Reagan & David Bowie\\
  \textbf{sv}& George W. Bush & Winston Churchill & Elizabeth II & Michael Jackson & Adolf Hitler\\
  \textbf{pl}& Elizabeth II & Pope John Paul II & Margaret Thatcher & George W. Bush & Ronald Reagan\\
  \textbf{fi}& Barack Obama & Adolf Hitler & Michael Jackson & George W. Bush & Benito Mussolini\\
  \textbf{no}& Marilyn Monroe & Adolf Hitler & John F. Kennedy & Bob Dylan & Bill Clinton\\
  \textbf{ru}& William Shakespeare & Napoleon II & Kenneth Branagh & Elton John & Joseph Stalin\\
  \textbf{zh}& Chiang Kai-Shek & William Shakespeare & Barack Obama & Deng Xiaoping & Adolf Hitler\\
  \textbf{ca}& Adolf Hitler & Che Guevara & Juan Carlos I & Michael Schumacher & Juan Manuel Fangio\\
  \hline
\end{tabular}
\end{table*}

\subsection{Most central persons}\label{sec:hist_rankings}
In this section we focus on centrality metrics for the above defined
biographical networks.  In Table~\ref{tab:top_indegree} we present the
top-ranked persons according to the degree centrality for the English
Wikipedia. We also provide results for other centrality measures
together with the corresponding rankings. Betweenness measures the
fraction of shortest paths between other pairs of nodes passing through
a given node, while PageRank gives a measure of the global
importance of nodes, computed recursively putting a larger weight on
incoming connections from central nodes.

We find many American presidents, iconic American musicians, and
European leaders during the WW2 period among the most
linked. Interestingly, we observe that Pope John Paul~II appears to be
a more central figure than Jesus.

Comparing the number of incoming and outgoing links, we observe that
in-degrees are of an order of magnitude greater than out-degrees. We
explain this phenomenon again by the presence of the parasocial
relations. For betweenness and PageRank we do not find large
differences in the rankings. The only exception is Shakespeare, whose
low betweenness value can be caused by the low number of out-going
links. Interestingly, Shakespeare's page is one of the most
central for several languages (see Table~\ref{tab:top_betweenness}),
but not for English.

In Table~\ref{tab:top_betweenness} we show the most central characters
in Wikipedia for the 15 analysed languages ranked by the betweenness
centrality.  We observe that most of the presented persons are known
to be (or have been) highly influential in many aspects. Thus, in
these lists we find political leaders, revolutionaries, famous
musicians, writers and actors.  We note that political figures such as
Adolf Hitler, George W. Bush or Barack Obama dominate in almost all
top rankings. Interestingly, William Shakespeare and Michael Jackson
are also among the central figures 
for several languages. 

For many languages we find, however, that the top ranked persons
reflect country specific issues.  Thus, for example, Pope John Paul II
is only present in the top five list of the Italian and Polish
Wikipedia, two countries which have a special tie with this figure.
In the English Wikipedia the most central figures are former US
presidents, while the Spanish-speaking Wikipedia community marks out
Latin American revolutionaries.  The Russian version surprisingly
highlights William Shakespeare and also Kenneth Branagh, known
for several film adaptations of Shakespeare's plays.  
Only the Japanese Wikipedia rankes the author Mishima
prominently, while two Chinese leaders, Chiang Kai-Shek and Deng
Xiaopin, are in the top-5 in the Chinese version.

When looking at these results, it should be taken into
account that there is an Anglo-Saxon bias in the dataset, as we relied
on a list of notable persons extracted from the English Wikipedia, and
persons from other cultures not know internationally might be
missing. In that sense the above list reflects centrality among
persons with at least limited international notoriety.

\begin{table}[!t]
  \caption{Similarities between the biographical networks of different 
    language Wikipedias.}
  \label{tab:jaccard}
  %\tiny
  \scriptsize{
  \begin{tabular}{@{}c@{\hspace{4pt}}c@{\hspace{4pt}}c@{\hspace{4pt}}c@{\hspace{4pt}}c@{\hspace{4pt}}c@{\hspace{4pt}}c@{\hspace{4pt}}c@{\hspace{4pt}}
  c@{\hspace{4pt}}c@{\hspace{4pt}}c@{\hspace{4pt}}c@{\hspace{4pt}}c@{\hspace{4pt}}c@{\hspace{4pt}}c@{\hspace{4pt}}
  c@{\hspace{4pt}}c@{\hspace{4pt}}c@{}}
\hline
& ca & de & en & es & fi & fr & it & ja & nl & no & pl & pt & ru & sv & zh\\
\hline
ca & - & .05 & .03 & \textbf{\underline{.12}} & .09 & .07 & .08 & .07 & .10 & .08 & .06 & \textbf{.10} & .06 & .09 & .06\\
de & .05 & - & .11 & .11 & .07 & \textbf{\underline{.13}} &  \textbf{.12} & .08 & .09 & .06 & .06 & .08 & .04 & .08 & .03\\
en & .03 & \textbf{\underline{.11}} & - & .09 & .03 &  \textbf{.10} & .08 & .05 & .05 & .03 & .03 & .05 & .02 & .04 & .02\\
es & .12 & .11 & .09 & - & .09 & .13 & \textbf{\underline{.14}} & .10 & .12 & .07 & .07 &  \textbf{.14} & .06 & .09 & .05\\
fi & .09 & .07 & .03 & .09 & - & .06 & .08 & .09 &  \textbf{.11} & .10 & .09 & .10 & .07 & \textbf{\underline{.13}} & .06\\
fr & .07 & .13 & .10 &  \textbf{.13} & .06 & - & \textbf{\underline{.15}} & .08 & .09 & .06 & .06 & .09 & .04 & .07 & .03\\
it & .08 & .12 & .08 &  \textbf{.14} & .08 & \textbf{\underline{.15}} & - & .09 & .10 & .07 & .07 & .11 & .05 & .08 & .04\\
ja & .07 & .08 & .05 & \textbf{\underline{.10}} & .09 & .08 & .09 & - &  \textbf{.10} & .08 & .07 & .09 & .05 & .09 & .08\\
nl & .10 & .09 & .05 &  \textbf{.12} & .11 & .09 & .10 & .10 & - & .10 & .09 & \textbf{\underline{.13}} & .07 & .12 & .05\\
no & .08 & .06 & .03 & .07 &  \textbf{.10} & .06 & .07 & .08 & .10 & - & .08 & .09 & .05 & \textbf{\underline{.13}} & .06\\
pl & .06 & .06 & .03 & .07 & .09 & .06 & .07 & .07 & \textbf{\underline{.09}} & .08 & - & .09 & .08 &  \textbf{.09} & .05\\
pt & .10 & .08 & .05 & \textbf{\underline{.14}} & .10 & .09 & .11 & .09 &  \textbf{.13} & .09 & .09 & - & .07 & .11 & .06\\
ru & .06 & .04 & .02 & .06 &  \textbf{.07} & .04 & .05 & .05 & .07 & .05 & \textbf{\underline{.08}} & .07 & - & .06 & .05\\
sv & .09 & .08 & .04 & .09 &  \textbf{.13} & .07 & .08 & .09 & .12 & \textbf{\underline{.13}} & .09 & .11 & .06 & - & .06\\
zh & .06 & .03 & .02 & .05 &  \textbf{.06} & .03 & .04 & \textbf{\underline{.08}} & .05 & .06 & .05 & .06 & .05 & .06 & -\\
\hline
  \end{tabular}
  }
\end{table}

\begin{figure}[!tb]
  \centering
  \includegraphics[width=\columnwidth]{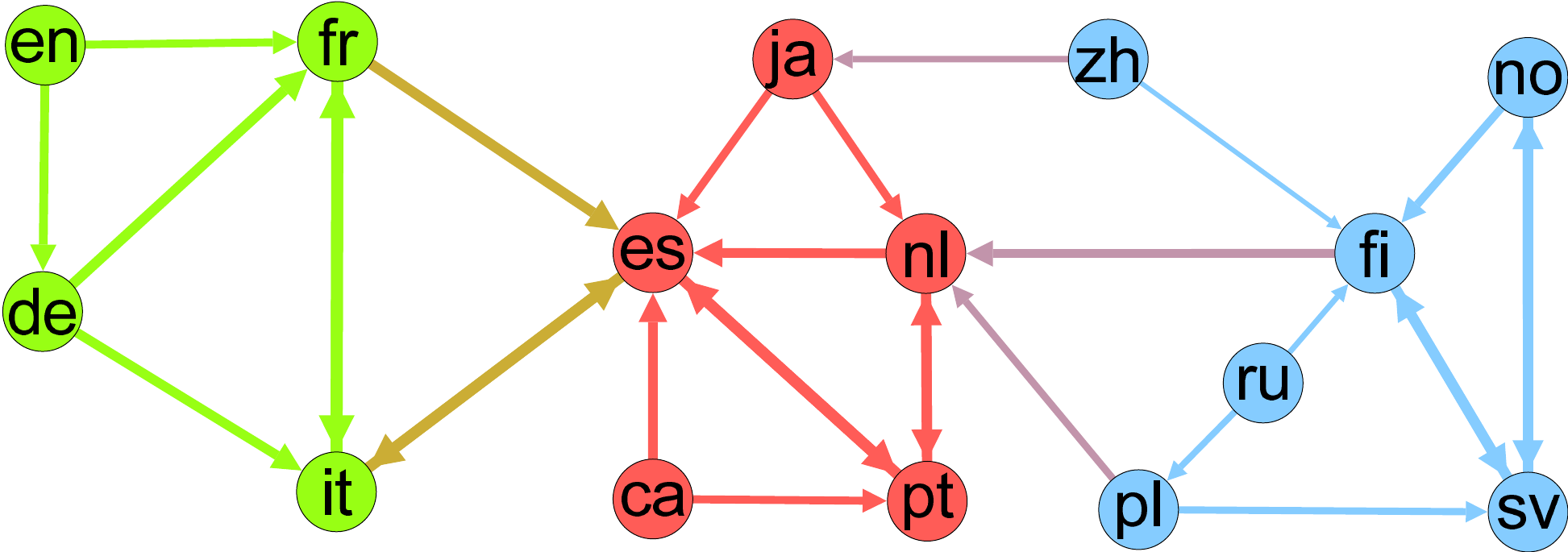}
  \caption{ Languages similarity network: every language links to the two languages with the largest similarities according to
    Table~\ref{tab:jaccard}.}
  \label{fig:jaccard_network}
\end{figure}

\begin{figure}[!tb]
  \centering
  \includegraphics[width=.5\textwidth]{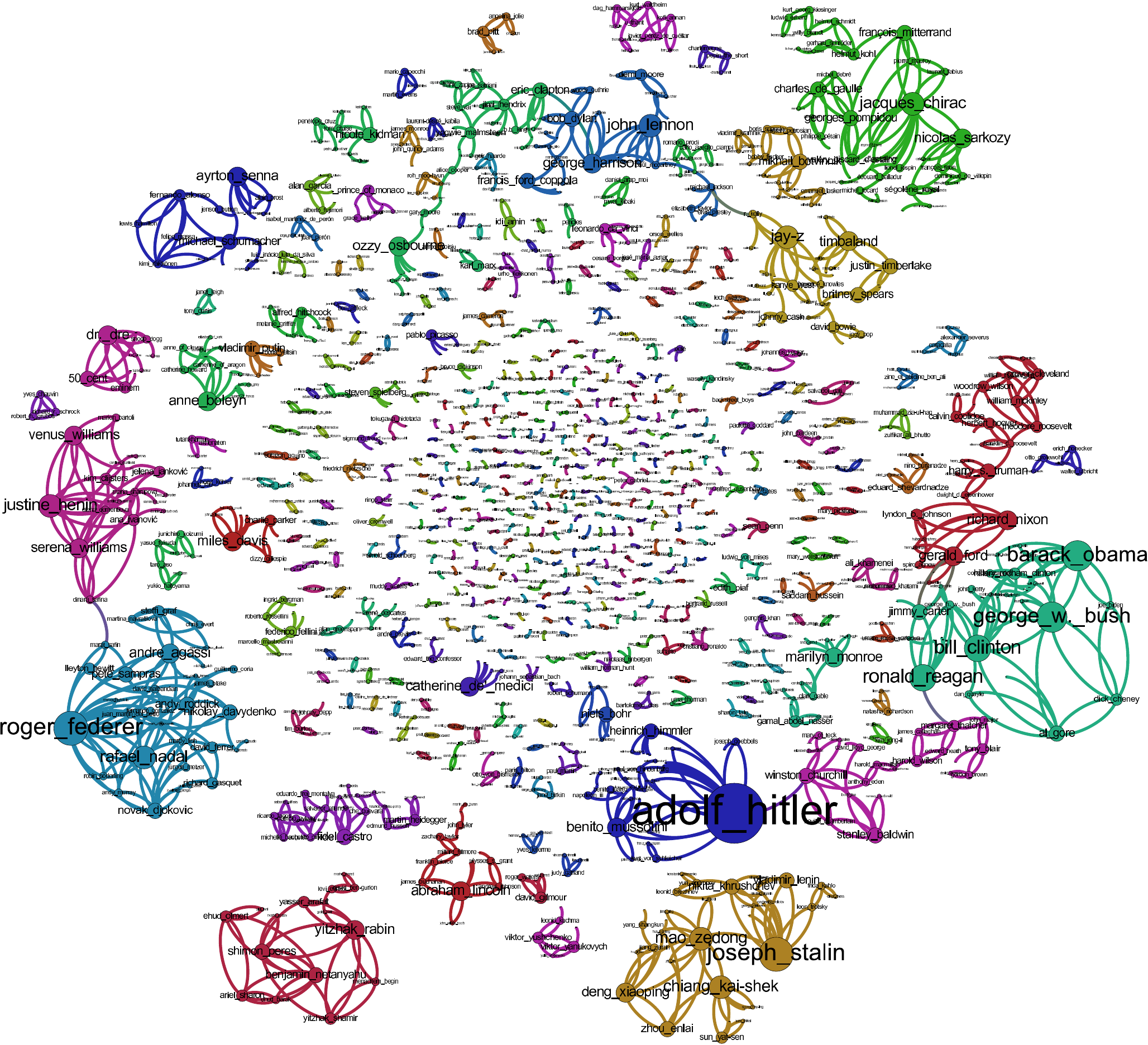}
  \caption{Biographical network of the connections in at least 13 of the 15
    analysed language versions. Larger version available at \url{http://goo.gl/Lc2Xl}.}
%\url{http://produccionmultimedia.barcelonamedia.org/var/www/public_html/wp-content/uploads/2012/04/intersection13_boundedV2.pdf}}
  \label{fig:the_network}
\end{figure}

\subsection{Similarity between languages}\label{sec_hist_languages}
In this section we focus on similarities between the networks emerging from %%%david: it was: %social networks of 
the different language Wikipedias. 
We calculate the similarity for every pair of networks as their Jaccard 
coefficient, i.e. the ratio between the
number of links present in both networks (their intersection) and
the number of links existing in their union.  
Table~\ref{tab:jaccard}
shows the obtained similarity results. For every language row we
highlight the two languages with the largest similarities. The most
similar language is also underlined.

{
Figure~\ref{fig:jaccard_network} further illustrates these
similarities by drawing a language similarity network. In this network
a language $A$ is connected to another language $B$ if language $B$ is
one of the two most similar to $A$. Applying the Louvain method, we
divide this graph into three clusters.  
In agreement with~\cite{liao2011analysing}, we observe that most of the
links can be explained by language-family relations (e.g Romance and
Slavic languages) and geographic or historical ties (e.g. Scandinavian
group, or Russia and Finland).  
We also find a number of less obvious
connections, e.g. Japanese to Spanish and Finnish to Dutch.  In fact,
Dutch seems to serve as a bridge between different language and
culture groups.
\sloppy

}
\subsection{Links present in most language
  networks}\label{sec_hist_union}

Finally, Figure~\ref{fig:the_network} depicts the network of
connections which are present in at least 13 of the 15 language
Wikipedias\footnote{The choice of 13 is arbitrary and taken mainly for
  visualization reasons.}. The network is relatively small, containing
1663 nodes and 1738 links, but allows to visualise the quintessence of
knowledge about biographical connections present in most of the
analysed language communities.

The largest connected component in Figure~\ref{fig:the_network}
corresponds to a cluster of US Presidents which connects over Ronald
Reagan to a cluster of British Premier Ministers. This group is
related through Winston Churchill to a cluster of persons from WW2's
axis powers.  The second largest component is compound of several
clusters related to the music and entertainment business, and the
third one of two clusters of male and female tennis players connected
through Dinara and Marat Safin.  Other large isolated clusters can be
found around such diverse groups as Russian and Chinese political
figures, French presidents, Israeli and Palestinian politicians,
Formula One pilots, World Chess Champions or actresses.

\section{Conclusions}\label{sec:conclusion}

Our results show that biographical connections are record\-ed differently
in the distinct language versions of Wikipedia. Although the global
social network measures are largely similar for all these networks,
the most central persons unveil interesting peculiarities about the
language communities.  A study of similarity
%among the networks
reveals that networks are more similar for geographically or
linguistically closer communities.
Nevertheless, there also exists a great number of biographical
connections which can be found in most of the analysed language
Wikipedias. Knowledge about these social connections trespasses
cultural barriers and represents part of the shared global collective
memory of our civilisation.

Possible directions for future work include the application of the
methodology to generate subnetworks of other kinds of article
categories, and a specific study of links present in most networks but
missing only in a few language Wikipedias.

Finally, the gender gap among Wikipedia editors is a serious 
concern for the community, and has been related %by previous work 
to the 
topics covered in the encyclopedia~\cite{laniado12emotions}. %\cite{lam_2011_clubhouse}. 
Our results point out a very small presence of females also among 
the most central persons in the encyclopedic content, suggesting the
link between these two phenomena as an intriguing subject for future 
investigation.

%ACKNOWLEDGEMENTS are optional
% \spara{Acknowledgements}

\section*{Acknowledgements} \small
 This work was partially supported by the Spanish Centre for the
 Development of Industrial Technology under the CENIT program, project
 CEN-20101037, “Social Media”. Yana Volkovich acknowledges support from
 the Torres Quevedo Program from the Spanish Ministry of Science and
 Innovation, co-funded by the European Social Fund.
\sloppy{

}

%\bibliographystyle{abbrv}
%\vspace{10pt}
%\small
%\bibliography{biblio}

\begin{thebibliography}{10}
\vspace{3mm}
\small

\bibitem{Alberich2002}
R.~Alberich, J.~Miro-Julia, and F.~Rossello.
\newblock {Marvel Universe looks almost like a real social network}.
\newblock {\em cond-mat/0202174}, 2002.

\bibitem{athenikos2009wikiphil}
S.~Athenikos and X.~Lin.
\newblock {The WikiPhil Portal: visualizing meaningful philosophical
  connections}.
\newblock {\em J. of the Chicago Colloq. on Digital Humanities and Comp. Sci.},
  1(1), 2009.

\bibitem{bellomi2005network}
F.~Bellomi and R.~Bonato.
\newblock Network analysis for {W}ikipedia.
\newblock In {\em Proc. of Wikimania}, 2005.

\bibitem{biuk2006visualizing}
R.~Biuk-Aghai.
\newblock Visualizing co-authorship networks in online {W}ikipedia.
\newblock In {\em Proc. of ISCIT'06}, 2006.

\bibitem{Chiyo2012}
P.~I. Chiyo, C.~J. Moss, and S.~C. Alberts.
\newblock The influence of life history milestones and association networks on
  crop-raiding behavior in male african elephants.
\newblock {\em PLoS ONE}, 7(2):e31382, 2012.

\bibitem{horton1956mass}
D.~Horton and R.~Wohl.
\newblock Mass communication and para-social interaction: Observations on
  intimacy at a distance.
\newblock {\em Psychiatry}, 19(3):215--229, 1956.

% \bibitem{lam_2011_clubhouse}
% S.~K. Lam, A.~Uduwage, Z.~Dong, S.~Sen, D.~R. Musicant, L.~Terveen, and
%   J.~Riedl.
% \newblock {WP}:clubhouse? an exploration of {W}ikipedia's gender imbalance.
% \newblock In {\em Proc. of WikiSym}, 2011.

\bibitem{laniado12emotions}
D.~Laniado, C.~Castillo, A.~Kaltenbrunner and M.~Fuster-Morell.
\newblock Emotions and dialogue in a large peer-production community: the case of {W}ikipedia.
\newblock In {\em Proc. of WikiSym}, 2012.

\bibitem{Laniado11coauthorship}
D.~Laniado and R.~Tasso.
\newblock Co-authorship 2.0: Patterns of collaboration in {W}ikipedia.
\newblock In {\em Proc. of Hypertext}, 2011.

\bibitem{Laniado2011}
D.~Laniado, R.~Tasso, Y.~Volkovich, and A.~Kaltenbrunner.
\newblock {When the {W}ikipedians talk: Network and tree structure of {W}ikipedia
  discussion pages.}
\newblock In {\em Proc. of ICWSM}, 2011.

\bibitem{liao2011analysing}
H.~Liao and T.~Petzold.
\newblock Analysing geo-linguistic dynamics of the world wide web: The use of
  cartograms and network analysis to understand linguistic development in
  {W}ikipedia.
\newblock {\em Cultural Science}, 3(2), 2011.

\bibitem{Mislove2007}
A.~Mislove, M.~Marcon, K.~P. Gummadi, P.~Druschel, and B.~Bhattacharjee.
\newblock Measurement and analysis of online social networks.
\newblock In {\em Proc. of IMC}, 2007.

\bibitem{pentzold2009fixing}
C.~Pentzold.
\newblock Fixing the floating gap: The online encyclopaedia {W}ikipedia as a
  global memory place.
\newblock {\em Memory Studies}, 2(2):255--272, 2009.

\bibitem{pfeil2006cultural}
U.~Pfeil, P.~Zaphiris, and C.~Ang.
\newblock Cultural differences in collaborative authoring of {W}ikipedia.
\newblock {\em JCMC}, 12(1):88--113, 2006.

\bibitem{zlatic2006wikipedias}
V.~Zlati{\'c}, M.~Bo{\v{z}}i{\v{c}}evi{\'c}, H.~{\v{S}}tefan{\v{c}}i{\'c}, and
  M.~Domazet.
\newblock Wikipedias: Collaborative web-based encyclopedias as complex
  networks.
\newblock {\em Physical Review E}, 74(1):016115, 2006.

\end{thebibliography}
% \appendix

\end{document}